\documentclass[prb,twocolumn,amsmath,amssymb,superscriptaddress,floatfix,aps]{revtex4-2}
\usepackage{graphicx,subfigure,amsmath,bm,colordvi,times}
\usepackage{mathtools}
\usepackage[usenames]{xcolor}
\usepackage{multirow}
\usepackage[utf8]{inputenc}
\usepackage[english]{babel}
\usepackage{amsfonts}
\usepackage{amssymb}
\usepackage{float}
\usepackage{hyperref}
\usepackage{braket}
\usepackage[normalem]{ulem}
\usepackage{comment}

\hypersetup{pdfborder=0 0 0,colorlinks=true,citecolor=blue,linkcolor=blue}

\begin{document}

\title{A Gross-Pitaevskii theory for an excitonic incompressible Bose solid}

\author{Sara Conti}
\affiliation{COMMIT, Department of Physics, University of Antwerp, Antwerp, Belgium}
\affiliation{TQC, Department of Physics, University of Antwerp, Antwerp, Belgium}

\author{Andrey Chaves}
\affiliation{Departamento de F\'isica, Universidade Federal do Cear\'a, Fortaleza, Cear\'a, Brazil}

\author{Alexander R. Hamilton}
\affiliation{School of Physics, University of New South Wales, Sydney, Australia}

\author{Jacques Tempere}
\affiliation{TQC, Department of Physics, University of Antwerp, Antwerp, Belgium}

\author{Milorad V. Milo\v{s}evi\'c}
\email{milorad.milosevic@uantwerpen.be}
\affiliation{COMMIT, Department of Physics, University of Antwerp, Antwerp, Belgium}

\author{David Neilson}
\email{dneilson@ftml.net}
\affiliation{COMMIT, Department of Physics, University of Antwerp, Antwerp, Belgium}


\begin{abstract}
We show that interlayer excitons in double-layer semiconductor heterostructures can form a Bose solid which is an incompressible supersolid characterized by exactly one boson per lattice site. 
This exciton Bose solid would be the first realization of an incompressible supersolid, unlike the generally compressible cluster supersolids seen in dipolar quantum gases.
Capturing its characteristics and associated emergent phenomena requires extending the Gross-Pitaevskii formalism to include strong two-particle correlations and exclude exciton self-interactions.
We develop such a formalism, we apply it across experimentally accessible exciton densities and interlayer separations, and we show that it incorporates both superfluid and incompressible supersolid ground states. 
This extended framework allows us to determine the superfluid–supersolid transition and explore the low-temperature properties of the exciton supersolid across its complete parameter space.
\end{abstract}

\maketitle


For many years there has been considerable theoretical interest in the Bose solid, but it has never been definitively observed experimentally.  
A true Bose solid must have a lattice site occupancy of exactly one boson per site~\cite{Anderson1984}. 
Precisely one boson per site generates a particle-antiparticle gap in the energy spectrum, and as a consequence, such a Bose solid will be incompressible.
References \cite{Anderson2012, Anderson2013} showed that the ground state of a Bose solid is an incompressible supersolid with the particles localized on their own lattice sites by strong particle-particle interactions, while at the same time they are in a single quantum condensate.
The Bose solid/incompressible supersolid is quite different from the superfluids with periodic clustering \cite{Bottcher2019, Natale2019}
or density modulation \cite{Fil2021} in dipolar cold atom systems that have been termed supersolid-like.
We have proposed the existence of such an incompressible supersolid in an exciton system based on a variational calculation \cite{Conti2023}.  

In this letter, we generalize the Gross-Pitaevskii approach to map out the ground state of the Bose solid at zero temperature. 
The Gross-Pitaevskii formalism is mean-field Hartree, so it includes interactions of a particle with itself~\cite{Gross1961, Pitaevskii1961}. 
However in a Bose solid with only a single particle on each site, there should be no self-interaction terms~\cite{Anderson2005}. 
Here, we extend the standard Gross-Pitaevskii approach to capture this property.
Furthermore, exchanges of bosons in the Bose solid may take place with difficulty because of the presence of other neighbors \cite{Anderson1984}.
We therefore also incorporate in our Gross-Pitaevskii formalism the effect of strong short-range two-particle correlations.
Development of a Gross-Pitaevskii based approach is a crucial enabler for further studies of static and dynamic phenomena in exciton superfluids and supersolids.

Our results show that excitons in a double-layer semiconductor heterostructure can form a Bose solid and would constitute the first experimental realization of an incompressible supersolid.
In this system, the excitons are spatially indirect, formed of electrons and holes that are confined in two adjacent two-dimensional conducting layers separated by a thin insulating barrier \cite{Burg2018,Ma2021}. 
With equal electron and hole densities, the electric field from the gates aligns the exciton dipoles perpendicular to the layers so they cannot deform or rotate. This ensures the dipole-like exciton interactions always purely repulsive, in contrast to free dipoles.  
Sufficiently strong repulsion can localize single excitons on separate lattice sites of a dipolar crystal \cite{Astrakharchik2007,Boning2011} similarly to electrons in a Wigner crystal \cite{Tanatar1989}. 
In both cases the purely repulsive interaction ensures only one particle per site.
This differs from cold atom systems where attractive as well as repulsive atomic interactions acting in competition are needed to generate periodic clusters \cite{Tanzi2019} which in general are not incompressible.


We start with the Hamiltonian for the excitonic state, 
\begin{eqnarray}
\widehat{H}&=& \int d^2\mathbf{r}\,  \widehat{\psi}^\dagger(\mathbf{r}) \left(-\frac{\hbar^2\nabla^2}{2M_{\text{X}}}\right)\widehat{\psi}(\mathbf{r})\\ 
&+& \frac{1}{2}\!\! \iint  d^2\mathbf{r} d^2\mathbf{r}' \, \widehat{\psi}^\dagger(\mathbf{r}) \widehat{\psi}^\dagger(\mathbf{r}')V_{\text{XX}}(|\mathbf{r}'\!\!-\!\mathbf{r}|) \widehat{\psi}(\mathbf{r}) \widehat{\psi}(\mathbf{r}'),
\nonumber
\label{Eq:H}
\end{eqnarray}
where $\widehat{\psi}(\mathbf{r})$ ($\widehat{\psi}^\dagger(\mathbf{r})$) 
is the exciton boson destruction (creation) operator at position $\mathbf{r}$ for the order parameter of the quantum condensate, and $M_{\text{X}}=m_e^*+m_h^*$ is the exciton mass.  
We set the electron and hole masses equal, $m_e^*=m_h^*$.

The area of the confined system is fixed by gates. The gates also fix the density and the number of particles.
We take equal electron and hole densities, the exciton dipoles align perpendicular to the layers, and, so the repulsive exciton-exciton interactions are,
\begin{equation} 
V_{\text{XX}}(r) =\frac{2 e^2}{4 \pi \epsilon} \left(\frac{1}{r}-\frac{1}{\sqrt{r^2+d^2}} \right)\ .
\label{Eq:V_XX}
\end{equation} 
Here $d$ is the distance between the layers,  $\epsilon$ is the dielectric constant of the insulating layer that separates the layers, and $\mathbf{r}$ is the separation of the excitons in the plane of the layers. 
$V_{\text{XX}}(r)$ is the sum of the four bare Coulomb interactions that act between the electrons and holes forming the two excitons.

At low densities, the $V_{\text{XX}}$ is well approximated by dipolar repulsion.   
From Quantum Monte Carlo (QMC) calculations, the pair correlation function $g(\mathbf{r})$ for dipoles at low densities vanishes for small $r$ \cite{Astrakharchik2007}. 
This effect is not considered in the conventional Hartree-Fock approach, but we may include it with an effective Hamiltonian ${\cal{H}}^{\rm{eff}}$ in which the effective exciton-exciton interaction contains a hard core of radius $R_{HC}$ that can be deduced from the structure factors taken from Ref.\ \cite{Astrakharchik2007}.
This keeps the excitons outside the region where the $g(\mathbf{r})$ as determined from QMC vanishes.  
It is equivalent to considering a local t-matrix $t_{\text{XX}}(\mathbf{r})$, representing multiple short-range two-body scattering, that vanishes for $r<R_{HC}$ \cite{Lowy1975}.
The effective Hamiltonian thus reads
\begin{eqnarray}
\widehat{{\cal{H}}}^{\rm{eff}} =\int &d^2\mathbf{r}&\,  \widehat{\psi}^\dagger(\mathbf{r}) 
\left(-\frac{\hbar^2\nabla^2}{2M_{\text{X}}}\right)\widehat{\psi}(\mathbf{r}) \nonumber\\ 
\!\!+  
\frac{1}{2}\! \iint\!  &d^2\mathbf{r}&\, d^2\mathbf{r}' \, \widehat{\psi}^\dagger(\mathbf{r}) \widehat{\psi}^\dagger(\mathbf{r}')t_{\text{XX}}(|\mathbf{r}'\!\!-\!\mathbf{r}|) \widehat{\psi}(\mathbf{r}) \widehat{\psi}(\mathbf{r}'), 
\label{Eq:Heff}
\end{eqnarray}
where
\begin{eqnarray}
t_{\text{XX}}(r)  &=& 
\left\{\begin{array}{ll}
V_{\text{XX}}(r) \ \ \  \mbox{  if } r>R_{HC}; \\
0 \ \ \ \ \ \ \ \ \ \ \ \ \ \mbox{otherwise. }
	\end{array}
 \right. \nonumber
\end{eqnarray}
For sufficiently strong exciton interactions, a normal-state exciton crystal can form, with exactly one exciton per lattice site \cite{Astrakharchik2007,Boning2011}.
This normal-state Bose solid can be described by a many-body wave function consisting of a product of single-particle wave functions. 
However in the absence of symmetry breaking, the solid many-body wave function will be translationally invariant, consisting of a superposition of an infinite number of infinitesimally translationally-shifted degenerate many-body states.  
The translation symmetry is broken by selecting one of these degenerate states, thereby making a choice of a particular set of sites $\{a_i\}$ in a fixed periodic lattice.  
The single-particle wave functions $\phi_i(\mathbf{r})$ will then be localized on the different sites $i$ of the lattice \cite{Anderson2012}. 
We derive a Hartree equation for this state in the supplemental material (SM) \cite{SupplementalGP}, as
\begin{eqnarray}
-\frac{\hbar^2\nabla^2}{2M_{\text{X}}} \phi_i(\mathbf{r})
 &+&  \int\!\!  d^2\mathbf{r}'  \phi_{i}(\mathbf{r})  
t_{\text{XX}}(|\mathbf{r}'\!\!-\!\mathbf{r}|) \sum_{\substack{j}} |\phi_j(\mathbf{r}') |^2  
\nonumber \\
&& \!\!\!\!\!\!\!\!\!\!\!\!\!\!\!\!\!\!\!\!\!\!\!\! 
- \int\!\!   d^2\mathbf{r}' \phi_{i}(\mathbf{r}) 
t_{\text{XX}}(|\mathbf{r}'\!\!-\!\mathbf{r}|) |\phi_i(\mathbf{r}') |^2 =\mu \phi_i(\mathbf{r}).
\label{Eq:minE_i} 
\end{eqnarray}

The effective attractive interaction term in Eq.\ \eqref{Eq:minE_i} is generated from the property that, while an exciton on a singly-occupied site $i$ feels the repulsive interactions from the excitons in the neighboring unit cells, it does not feel its own potential within the unit cell $i$ \cite{Anderson1984} (see Eq.\ (8) in the SM \cite{SupplementalGP}). 
This term removes the contributions in the Hartree equation from a single boson in a unit cell interacting with itself. 
The term is only present when there is exactly one particle per unit cell, since with more particles there would be interactions within the unit cell.
This term is not an externally imposed potential but self-consistently arises from the exciton-exciton interactions.

The effect of the absence of these self-interactions within a site acts like a potential well centered on the site.  
For sufficiently strong repulsion between the excitons, the potential wells can become deep enough for bound-state solutions to exist \cite{Anderson1984}. 
In fact, in systems with purely repulsive particle-particle interactions such as electrons or perpendicularly aligned excitons, this is the mechanism that can drive the solidification \cite{Anderson2012}, and in these repulsive systems quantum Monte Carlo simulations have confirmed the existence of normal solids \cite{Tanatar1989,Astrakharchik2007,Boning2011}. 
When there is a bound state solution, the missing self-interaction terms will open up an energy gap between the particle and anti-particle excitations \cite{Anderson2005}. 
The anti-particle excitations are then localized in the self-consistently created wells while the particle excitations with their positive energies are free-running waves. 
The resulting solid would be incompressible since the potential wells lead to a prohibitive energy cost against doubly occupied sites \cite{Anderson2005}, analogous to a Mott insulator.

In the present study we investigate the possibility of solidification coexisting with quantum coherence in a supersolid.
We extend the standard Gross-Pitaevskii formalism to include the effect of the absence of self-interactions specific to this exciton system.
In place of the many-body wave function of the normal state system, we construct the order parameter for the incompressible supersolid of the form, 
\begin{equation}
\Psi(\mathbf{r}) =  
 \sqrt{\rho_B} + \sum_{i=1}^{N} 
\phi_i (\mathbf{r}) \ ,
\label{Eq:Psi}
\end{equation}
where $i$ is the site index for the $N$ sites.  
With this construction, each exciton has the same wave function spread across the entire lattice and with peaks at every site.
This order parameter results in both off-diagonal long-range order and also diagonal long-range order, the latter coming from the lattice symmetry.
In addition, for a supersolid there must exist phase locking between adjacent sites.  This arises from mutual overlap of the $\phi_i(\mathbf{r})$ \cite{Leggett1970}. 
With the $\phi_i(\mathbf{r})$ exponentially localized on their sites it is not possible to link the phases on the different lattice sites, since zero-point quantum and thermal fluctuations eliminate coherence between the near-negligible overlaps.
We ensure significant overlap by means of the uniform background component $\rho_B$ in Eq.\ \eqref{Eq:Psi}. 
In cases when $\rho_B$ tends to zero there is negligible overlap, so there can be no supersolid. 
In cases when the $\phi_i(\mathbf{r}-\mathbf{r}_i)$ reduce to zero, $|\Psi(\mathbf{r})|^2 \sim \rho_B$, and Eq.\ \eqref{Eq:Psi} represents the homogeneous order parameter for a superfluid. 

From Eq.\ \eqref{Eq:minE_i}, we derive a time-independent Gross-Pitaevskii equation \cite{RogelSalazar2013} for the order parameter $\Psi(\mathbf{r})$ (see the SM \cite{SupplementalGP}),  
\begin{eqnarray}
&&\!\!\!\!\!\!\frac{\hbar^2\nabla^2}{2M_{\text{X}}}  \Psi(\mathbf{r}) + \! \int \!\! d^2 \mathbf{r}'\,  t_{\text{XX}}(|\mathbf{r}'\!-\!\mathbf{r}|)|\Psi(\mathbf{r}')|^2 \, \Psi(\mathbf{r}) \nonumber\\
&&-\!\!\int_{\mathbf{r}' \in i_\mathbf{r}}
\!\!\!\!\!\!  d^2\mathbf{r}' 
t_{\text{XX}}(|\mathbf{r}'\!-\!\mathbf{r}|) 
|\Psi(\mathbf{r}')|^2 (\Psi(\mathbf{r})\!-\!\sqrt{\rho_B}) =\mu \Psi(\mathbf{r})   . 
\label{Eq:GP_SS}
 \end{eqnarray}
The notation ${\mathbf{r}'\in i_\mathbf{r}}$ indicates that the integral over $\mathbf{r}'$ is restricted to the same $i_\mathbf{r}$ unit cell as $\mathbf{r}$.

\begin{figure*}[t]
\includegraphics[width=0.245\textwidth]{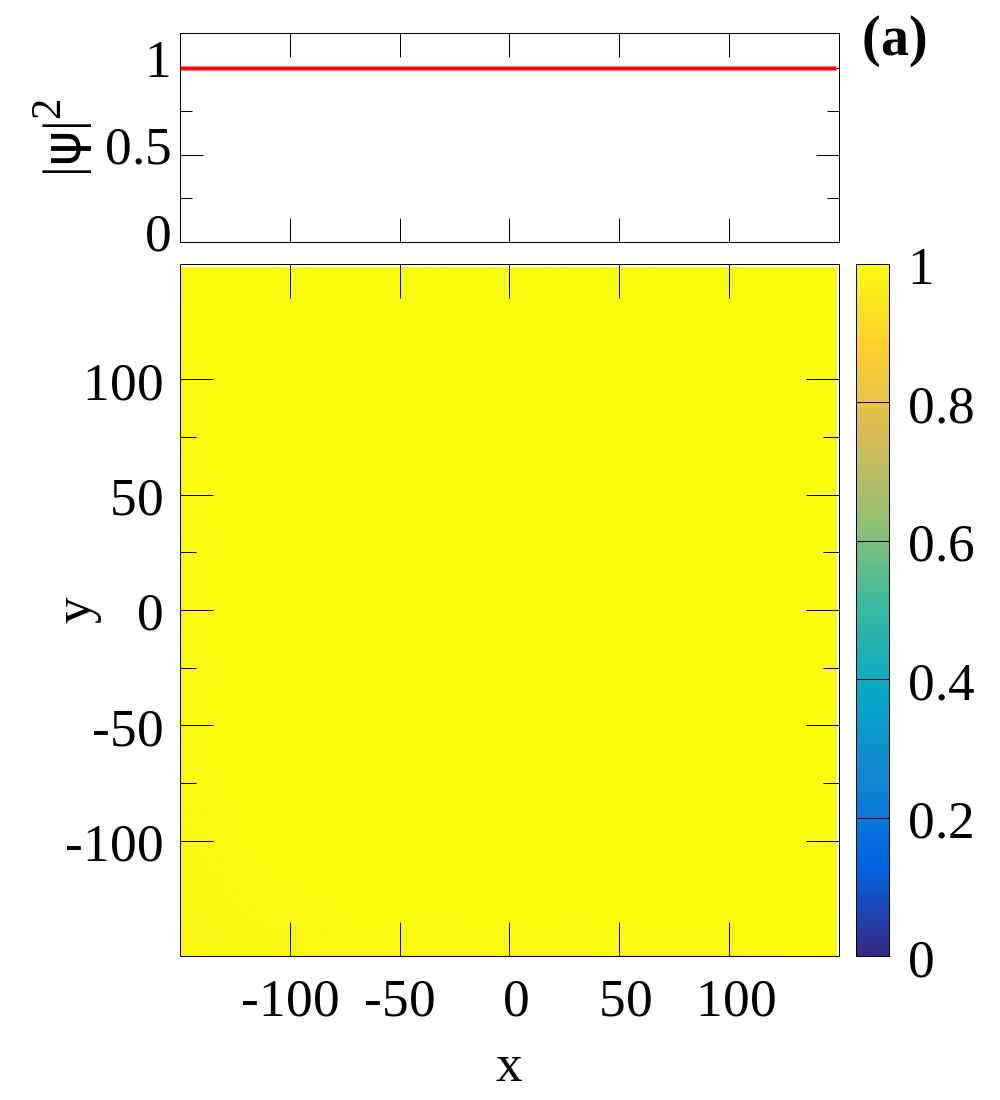}
\includegraphics[width=0.245\textwidth]{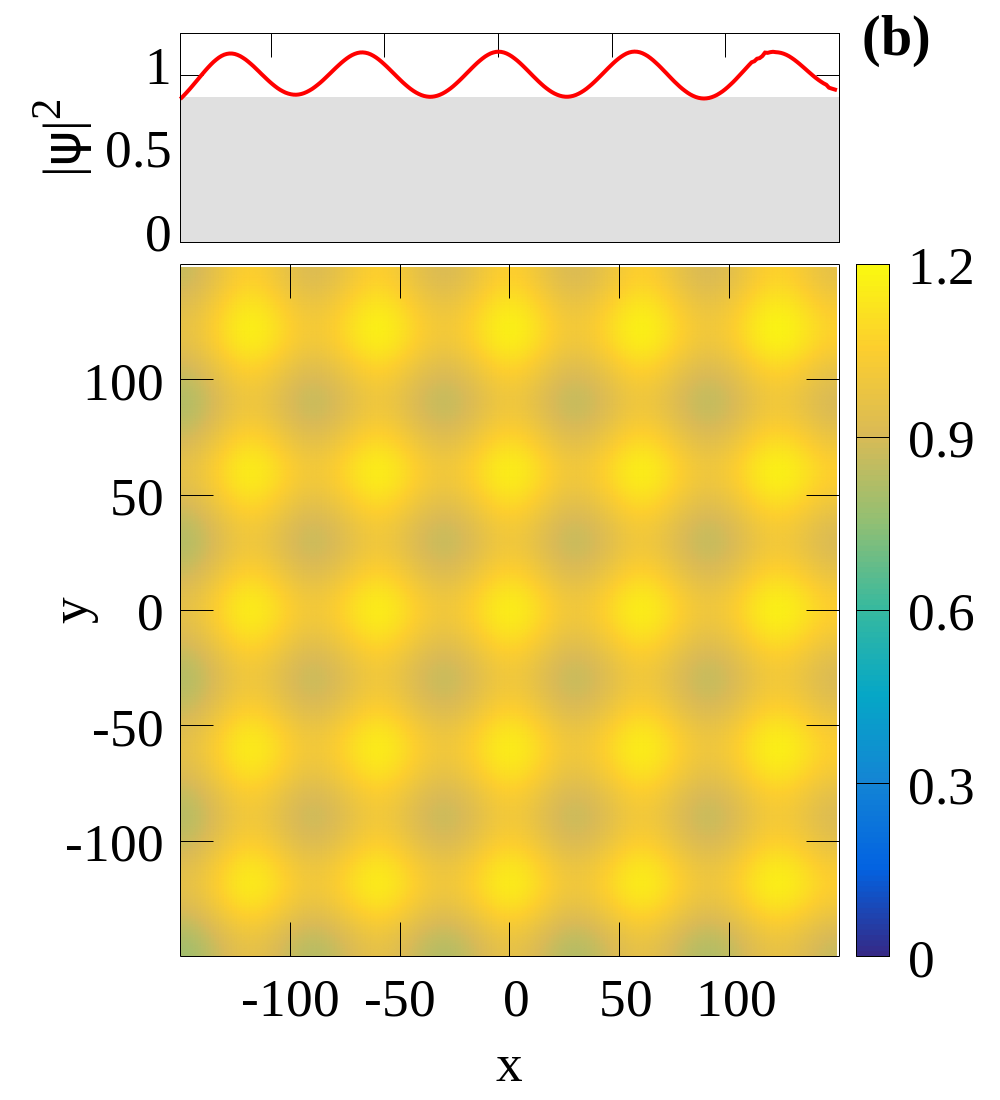}
\includegraphics[width=0.245\textwidth]{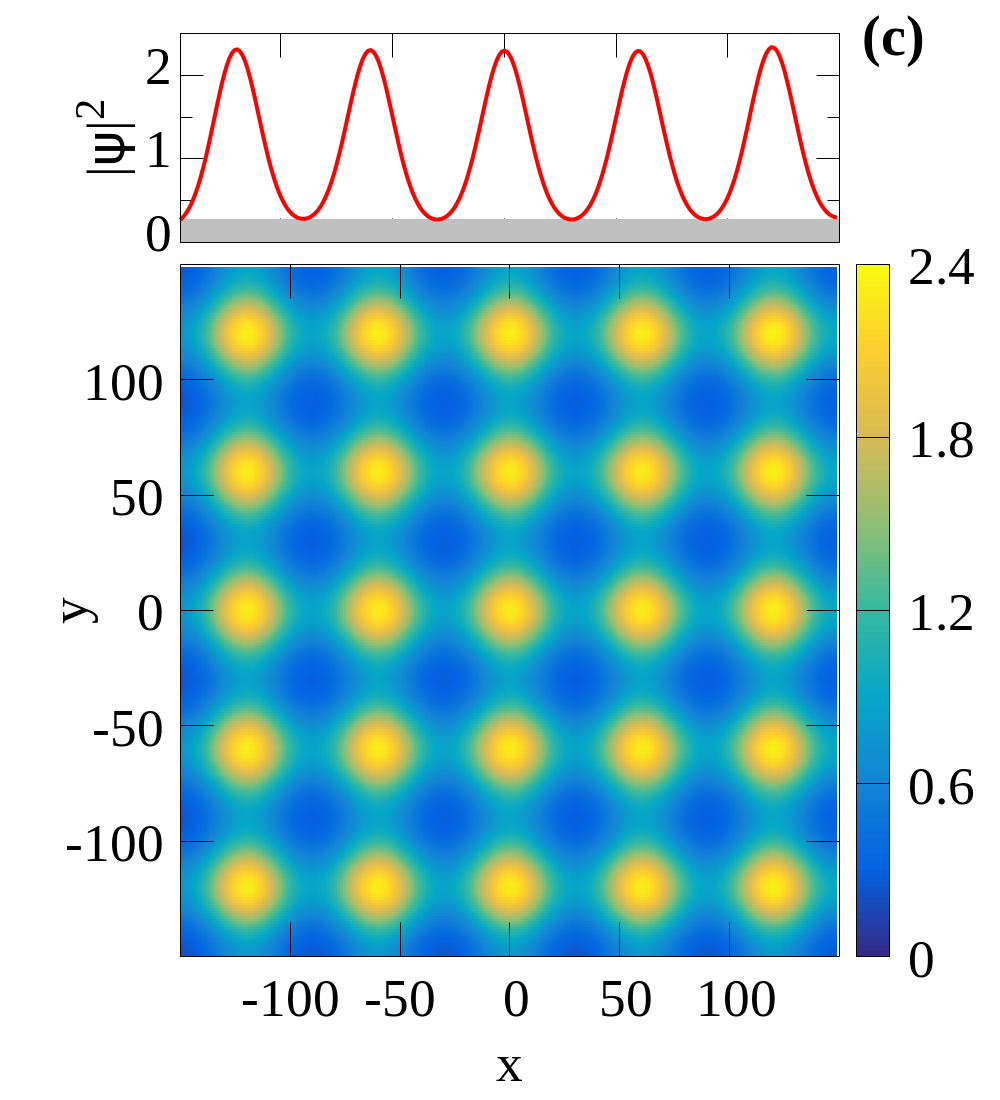}
\includegraphics[width=0.245\textwidth]{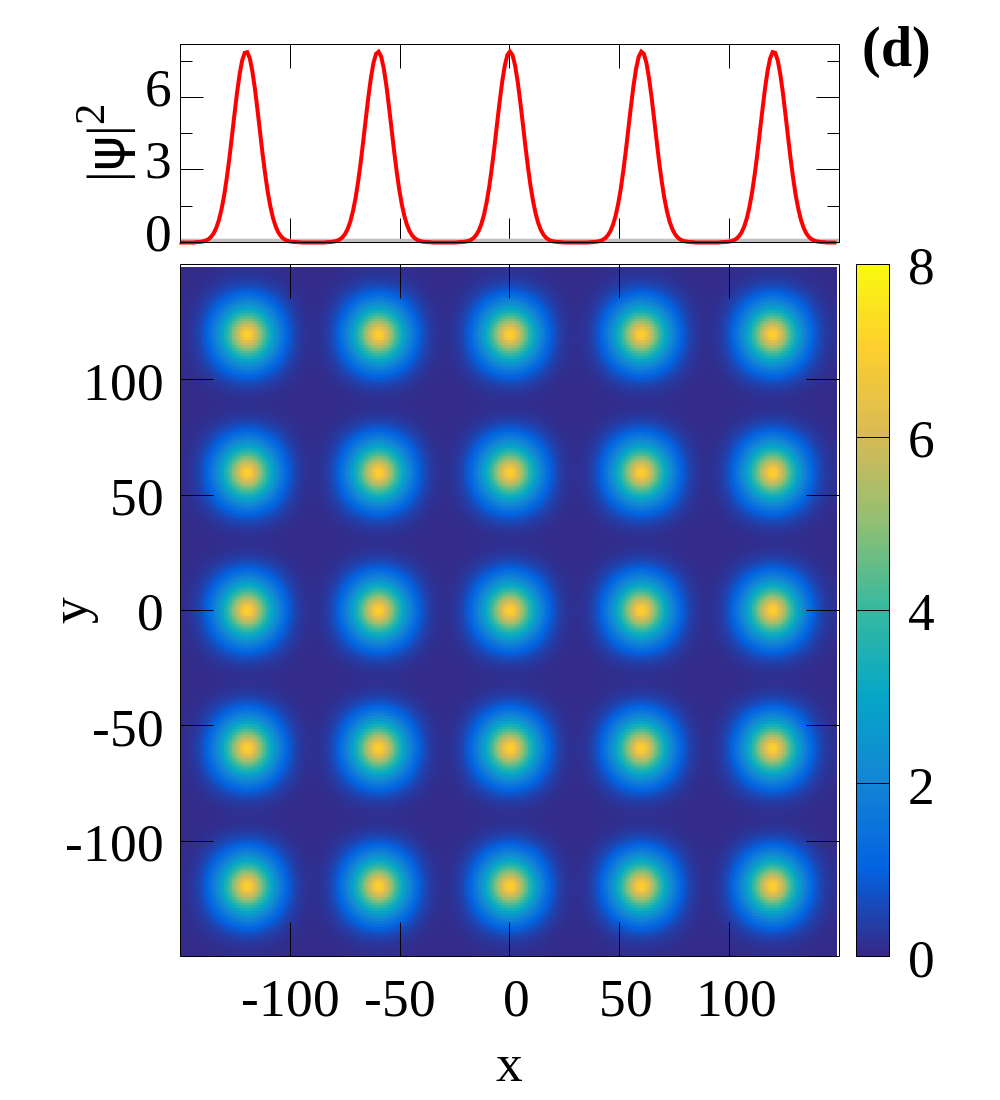}
\caption{Profile along diagonal and the spatial distribution of order parameter $|\Psi(\mathbf{r})|^2$  for $r_0=30$. 
Panels (a)--(d) correspond to the interlayer separations $d=1,7,15,20$, respectively. 
Shaded area shows the homogeneous background component of $|\Psi(\mathbf{r})|^2$.}
\label{fig:r0=30d=1}
\end{figure*}

The mean-field potential in Eq.\ \eqref{Eq:GP_SS} differs from the mean-field potential in the standard Gross-Pitaevskii equation for dipolar bosons \cite{Yi2001}, 
not only because $V_{\text{XX}}$ has been replaced by $t_{\text{XX}}$ to account for two-body correlations, but also because  there is an additional potential term. 
It is this term, that accounts for the absence of self-interactions and makes supersolid solutions possible.  


We solve Eq.\ \eqref{Eq:GP_SS} by evolving in imaginary time \cite{Haas2018}, starting from a spatially randomized order parameter. 
Note throughout the numerical calculation there is no restriction that $\Psi(\mathbf{r})$ is of the form of Eq.\ \eqref{Eq:Psi}. 
Because of the non-local interaction $V_{\text{XX}}(\mathbf{r})$, we evaluate the mean-field potential by means of the convolution theorem \cite{Fischer2006}.

Analogously to the normal solid, to identify the supersolid we fix the lattice positions $\{a_i\}$ 
thereby breaking the translational symmetry ourselves.  
We choose a square lattice even though in two-dimensions a triangular lattice could have a slightly lower energy.  
The triangular lattice is computationally very demanding and the issue we are interested in addressing here is the existence of a supersolid. 
In our static calculation, if the energy of the square lattice supersolid is less than the energy of the superfluid, then so will be the energy of the triangular lattice supersolid.
Since the excitons are gate confined within a fixed area of the layers, with one exciton per site the lattice constant $a$ for a square lattice is geometrically related to the exciton density $\rho=1/\pi r_0^2$ by $a=2r_0$.  

All simulations were performed on a $256^2$ grid covering the sample spatial area. 
Solutions were taken as converged when the relative change in energy in consecutive time steps decreases below $10^{-8}$.
After reaching convergence, we have verified for self-consistency that $\Psi(\mathbf{r})$ indeed has the form of Eq.\ \eqref{Eq:Psi}, with $\rho_B$ not zero.
We focus first on the low-density regime, setting the average interparticle spacing $r_0=30$. 
As length and energy scales, we use the effective Bohr radius $a_B=\hbar^2\epsilon/e^2 m_e^*$ and the effective Rydberg Ry$^*$.

Figure \ref{fig:r0=30d=1}(a) shows the ground-state order parameter $|\Psi(\mathbf{r})|^2$ for small interlayer separation $d=1$. 
The solution $\Psi(\mathbf{r})$ is homogeneous -- corresponding to a superfluid: for such a small $d$, the exciton dipole interactions  and the self-consistently calculated self-interaction potential are very weak compared with the kinetic energy, too weak to drive the system to a solid.

It is interesting to note that for this small value of $d$, the solution is unchanged if in Eq.\ \eqref{Eq:GP_SS} we do not include the negative term that removes the self-interaction. 
The self-interactions thus have no effect on the ground state for such weak dipolar interactions.

\begin{figure}[t]
    \centering    
    \includegraphics[trim={0.0cm 0.0cm 1.5cm 0.0cm},clip=true,width=0.9\columnwidth]{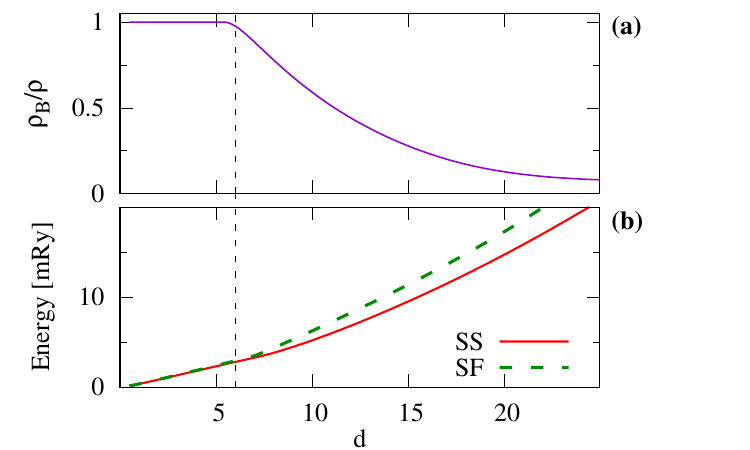}
    \caption{(a) Dependence on interlayer separation $d$ of $\rho_{B}/\rho$, the amplitude of the homogeneous background component of $\Psi(\mathbf{r})$, for $r_0=30$. 
 (b) Energy eigenvalue of the supersolid state (SS) and corresponding eigenvalue of the superfluid state (SF), calculated without the self-interaction correction.}
    \label{fig:Energies}
\end{figure}

By increasing $d$ we increase the dipole moment, making the exciton interactions stronger. 
In Fig.\ \ref{fig:r0=30d=1}(b) for $d=7$, we see that the order parameter $\Psi(\mathbf{r})$ is now inhomogeneous with a periodicity of the lattice.  
The $\Psi(\mathbf{r})$ has a normalized peak at each lattice site and there is a large overlap between adjacent sites.
Thus we conclude that this is an incompressible supersolid ground state, with the order parameter $\Psi(\mathbf{r})$ indeed of the parametrized form of Eq.\ \eqref{Eq:Psi}.
The overlap that allows exchange of bosons is associated with the homogeneous background component of amplitude $\rho_{B}$ (shaded area). 

Figures \ref{fig:r0=30d=1}(c) and \ref{fig:r0=30d=1}(d) show that when the interlayer spacing $d$ is further increased, the peak in $\Psi(\mathbf{r})$ on each site narrows and the amplitude $\rho_{B}$ of the homogeneous background component decreases. 
It may seem at first sight surprising that the lattice constant does not depend on $d$ and the effective interaction strength, but this is because the particle density is fixed by the gate charge and the repulsive interaction only allows one particle per unit cell (in a similar manner the lattice constant of a 2D Wigner crystal is independent of the strength of the Coulomb interaction). 

Figure \ref{fig:Energies}(a) shows the dependence of the amplitude $\rho_{B}$ on $d$ for fixed $r_0=30$. 
The behavior of $\rho_{B}$ is similar to that obtained in the variational calculation of Ref.\ \cite{Conti2023}.
$\rho_{B}/\rho=1$ corresponds to a homogeneous superfluid solution, while for a supersolid solution, $0<\rho_{B}/\rho<1$. 
The background component decreases with increasing $d$ but for a supersolid solution it does not vanish. 

Figure \ref{fig:Energies}(b) compares the supersolid energy eigenvalue $\mu_{SS}$ from Eq.\ \eqref{Eq:GP_SS}, and hence with the broken translational symmetry of the fixed lattice, compared with the superfluid energy eigenvalue $\mu_{SF}$ determined without the subtracted self-interaction term in Eq.\ \eqref{Eq:GP_SS}, and thus without the broken translational symmetry.  
For $d<6$ the two energy eigenvalues are identical, indicating a stable superfluid, but 
for $d\gtrsim 6$ the energy eigenvalue $\mu_{SF}$ is higher than $\mu_{SS}$ indicating that the incompressible supersolid becomes the ground state for $d\gtrsim 6$.

\begin{figure}[t]
    \centering    
    \includegraphics[width=1\columnwidth]{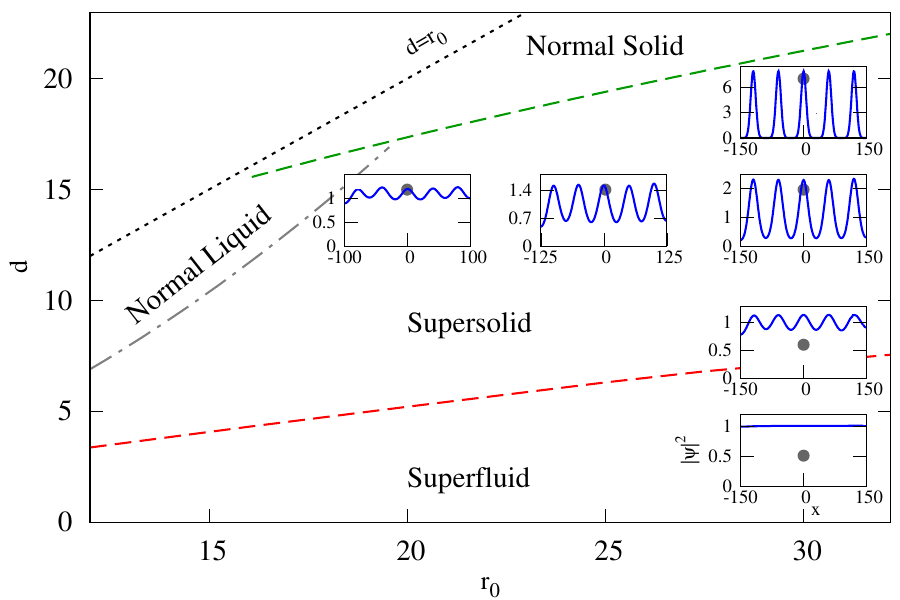}
\caption{The phase diagram at zero temperature, as a function of interlayer separation $d$ and average interparticle spacing $r_0$. 
Dashed red line: superfluid to supersolid transition.
Dash-dotted grey line: Superfluid and supersolid condensate collapse into a normal electron-hole liquid \cite{Pascucci2025note}.
Dashed green line: exciton liquid to normal exciton solid transition taken from Ref.\ \cite{Astrakharchik2007}.
Dotted black line: $d=r_0$.
Insets: order parameter profile $|\Psi(\mathbf{r})|^2$ at the points indicated by grey dots.
}
    \label{fig:PhaseDiagram}
\end{figure}

Repeating the above analysis for increasing density allows us to map out the $r_0$-$d$ phase diagram at zero temperature, shown in Figure \ref{fig:PhaseDiagram}.  
For a fixed low density ($r_0=30$) and small $d$, the exciton interaction $V_{\text{XX}}$ is weak, leading to a superfluid coherent phase with a constant order parameter $\Psi(\mathbf{r})$ (inset at $d=2.5$).  
When $d$ is increased, our Gross-Pitaevskii formalism predicts a transition from the superfluid to an incompressible supersolid (dashed red line).  
This lies very close to the corresponding transition line predicted from the variational calculation in Ref.\ \cite{Conti2023}.

At very large $d$ a normal exciton solid is expected to become the ground state \cite{Astrakharchik2007,Boning2011,Conti2023}. 
In contrast to the supersolid, in the normal exciton solid, the peaks of the wave function are independently localized without exchange.
Interestingly, our Gross-Pitaevskii approach does reflect an evolution towards the normal exciton solid for increasing $d$: the peaks in the order parameter become narrower and the amplitude of the homogeneous background trends towards zero (insets for $d=30$). 
However, the Gross-Pitaevskii formalism cannot determine the transition to the normal exciton solid since it \textit{a priori} assumes a global phase coherence of one order parameter with a non-zero homogeneous background.

Figure \ref{fig:PhaseDiagram} also shows the effect on the supersolid $\Psi(\mathbf{r})$ of increasing the density for a fixed $d$ (see insets at $d=15$).  
As $r_0$ decreases, the peaks in $\Psi(\mathbf{r})$ become broader and the homogeneous background component of $\Psi(\mathbf{r})$ grows larger.
This is associated with the evolution with $r_0$ of the average exciton interaction $V_{\text{XX}}$ (Eq.\ \eqref{Eq:V_XX}) from a dipolar form to a coulombic form. 
Coulomb interactions are known to support crystallization only at very low densities \cite{Swierkowski1993,DePalo2002}, and so decreasing $r_0$  has the effect of softening the supersolid. 
The supersolid phase shown in Fig.\ \ref{fig:PhaseDiagram} lies almost everywhere in the BEC regime, with condensate fractions of order unity, where screening is negligible or very weak \cite{Neilson2014}.  However, before the supersolid melts with decreasing $r_0$, the exciton condensate passes into the BEC-BCS crossover regime where screening would rapidly switch on \cite{Perali2013}, killing the quantum coherence. 
The transition from quantum coherence to the normal state is indicated by the grey dash-dotted line \cite{Pascucci2024,Pascucci2025note}.
In this high-density region of phase space, neither coulombic nor dipolar interactions support crystallization in the normal state \cite{Swierkowski1993,DePalo2002,Astrakharchik2007}.
Thus we expect that when the quantum coherence is lost, simultaneously the supersolid will melt to the normal-state liquid.

Since the BEC to BEC-BCS crossover boundary lies so close to the melting curve \cite{Pascucci2024}, our Gross-Pitaevskii formalism with bosonic single particles interacting without screening should be valid for essentially all of the area covered by the superfluid and supersolid phases shown in Fig.\ \ref{fig:PhaseDiagram}.

Realistic parameters for exploring the supersolid phase can be extracted from Fig.\ \ref{fig:PhaseDiagram} with knowledge of the effective Bohr radius $a_B$ for a specific system.  
As an example, in a TMD bilayer with an hBN barrier, $a_B\sim1$ nm, the supersolid phase is experimentally accessible with interlayer separations $d \sim 10-20$ nm and densities $\sim 1-5 \times 10^{11}$ cm$^{-2}$.

Reference \cite{Boning2011} shows that the Berezinskii-Kosterlitz-Thouless  temperature for the normal to superfluid transition and the melting temperature of the normal solid-liquid transition are both of the order of a few mRy$^*$. The supersolid melting temperature is also a Berezinskii-Kosterlitz-Thouless transition. Thus we expect the supersolid melting temperature to be of the same order, corresponding in the TMD bilayer to $\sim2$ K, again readily experimentally accessible.

In conclusion, we have demonstrated that an incompressible Bose solid/supersolid with one particle per site \cite{Anderson1984} should be experimentally accessible for a system of excitons in an electron-hole double-layer heterostructure. 
Our approach is based on the Gross-Pitaevskii formalism and can map out the quantum condensed phases over the zero-temperature phase diagram. 
Starting with the formalism for aligned bosonic excitons with purely repulsive dipolar-like interactions, we extended the formalism (i) to exclude the self-interaction energies, which should be absent for single occupancy sites \cite{Anderson2013}, and (ii) to take into account strong two-particle correlations. 

For weak exciton-exciton interactions our formalism predicts a superfluid ground state, but for stronger interactions there is a transition to an incompressible supersolid/Bose solid, with an order parameter that has peaks on a periodic lattice of sites superimposed on a uniform background.   
The background component found in our calculations is necessary for the phase coherence of the incompressible supersolid \cite{Leggett1970}. 
For even stronger interactions, our formalism displays an evolution towards a normal exciton solid, with the peaks in the order parameter narrowing and increasing in amplitude while its uniform background component decreases towards zero.  
In a normal exciton solid, the peaks would be independently localized without phase coherence between the peaks. 
In that respect, the evolution from supersolid to normal solid resembles that observed in the superfluid to Mott insulator transition in quantum gases \cite{Bloch2022}.

Within the supersolid phase, as the density is increased the exciton repulsive interaction (Eq.\ \eqref{Eq:V_XX}) evolves from dipolar-like to Coulomb-like, making the supersolid soften and eventually melt.  
 Just before melting, the system enters the BEC-BCS crossover regime where screening rapidly becomes strong and the system simultaneously loses coherence and crystal structure. 
The quantum condensed phases lie almost entirely within the BEC regime where screening is negligible, allowing us to describe the system in terms of a Bose condensate with a macroscopic order parameter. 
This makes our Gross-Pitaevskii formalism valid across the area of the phase diagram encompassing both the supersolid and superfluid phases.

To observe the exciton supersolid, both spatial periodic order and quantum coherence must be experimentally confirmed.  
For the spatial order, photoluminescence has been used to detect signatures of crystallization in gate-controlled exciton bilayers \cite{Bai2023} and optically pumped exciton bilayers \cite{Lagoin2024}.  Scanning Tunneling Microscopy (STM) has been used to directly image a 2D crystal lattice of electrons in a van der Waals heterostructure with sub nanometer resolution through a monolayer graphene gate \cite{Li2021}, and this technique should also be applicable to imaging the charge distribution of the top layer in a gated exciton bilayer.  
Existence of quantum coherence can be demonstrated using photoluminescence interference to detect off-diagonal long-range order \cite{Anankine2017, Cutshall2025}  and to detect vortices \cite{Anankine2017, Conti2025}.

Our Gross-Pitaevskii formalism establishes strong foundations for exploring the elusive incompressible supersolid and for identifying its hallmark features.
For example, by studying the formation of vortices and vortex lattices \cite{Klaus2022,Casotti2024,Conti2025} and extending the formalism to the time-dependent Gross-Pitaevskii equation, the dynamics of the vortices \cite{Gallemi2020,Ancilotto2021} can be used to identify the transition to an incompressible supersolid.
This powerful framework also gives access to the rich dynamics of phase transitions to investigate critical phenomena and scaling behavior, as well as access to the collective excitations of the condensate, capturing the behavior of the low-energy collective modes to provide insights into the thermodynamics of the Bose solid.

We thank Filippo Pascucci for very useful discussions.
This work was supported by Research Foundation -- Flanders (FWO-Vl), with contract number 12A3T24N; by the University Research Fund (BOF), with project numbers 	
53228, G0A9F25N, G0AIY25N, GOH1122N, G061820N, and G060820N;  by the Brazilian Council for Research (CNPq), through the PRONEX/FUNCAP, Universal, and PQ programs; and by the Australian Research Council projects CE170100039, DP210101608 and IL230100072. 


\begin{thebibliography}{40}
\bibitem [{Sup()}]{SupplementalGP}%
  \BibitemOpen
  \href@noop {} {}\bibinfo {note} {See {S}upplemental {M}aterial for derivation
  of the Gross-Pitaevskii equation, which includes {R}efs.\ [2], [13], [19] and
  [20]}\BibitemShut {NoStop}%
\end{thebibliography}
%

\clearpage
\onecolumngrid
\appendix
\section{Derivation of Gross-Pitaevskii equation for an exciton Bose solid}

We outline here the development of our generalized Gross-Pitaevskii formalism.

Consider the wave function for a Bose solid \cite{Anderson2012} in the normal state,
\begin{equation}
\ket{\Psi_{\mathrm{BS}}(\mathbf{r})} = \displaystyle \prod_{m=1}^N \hat{c}_m^\dagger \ket{\Psi_{\mathrm{vac}}(\mathbf{r})} \ .
\label{Eq:GroundState}
\end{equation}
$\ket{\Psi_{\mathrm{vac}}(\mathbf{r})}$ is the vacuum, 
$ \hat{c}_m^\dagger = \int d^2\mathbf{r} \, \phi_m^\star(\mathbf{r}) \widehat{\psi}^\dagger(\mathbf{r})$,
 $\hat{c}_m = \int d^2\mathbf{r} \, \phi_m(\mathbf{r}) \widehat{\psi}(\mathbf{r})$ with the $\phi_m(\mathbf{r})$ a set of  functions, identical but each centered 
on a different site $\{m\}$ of a periodic lattice. 
The $\phi_m(\mathbf{r})$ are degenerate and mutually orthogonal. 

The effective Hamiltonian (Eq.\ ($3$) in the main manuscript) can be written, 
\begin{equation}
 \widehat{{\cal{H}}}^{\rm{eff}}= \sum_{i,j} \hat{c}_i^\dagger T_{ij} \hat{c}_j +\frac{1}{2}\!\! \sum_{i,i',j,j'} \hat{c}_i^\dagger \hat{c}_{i'}^\dagger W_{i,i',j,j'} \hat{c}_j \hat{c}_{j'} \ ,
\label{Eq:Heffnew}
\end{equation}
with kinetic and potential energy ground state expectation values,
\begin{eqnarray}
T_{ij} &=&\int\,d^2\mathbf{r}\, \phi_i^\star(\mathbf{r}) \left(-\frac{\hbar^2\nabla^2}{2M_{\text{X}}} \right) 
 \phi_j(\mathbf{r})   \ ,
 \label{Eq:T}
  \\
W_{i,i',j,j'}\!\! &=& \!\!
\iint\!\!  d^2\mathbf{r}\, d^2\mathbf{r}'  \phi_i^\star(\mathbf{r}) \phi_{i'}^\star(\mathbf{r}') 
t_{XX}(|\mathbf{r}'\!\!-\!\mathbf{r}|) \phi_j(\mathbf{r}') \phi_{j'}(\mathbf{r})  \ .
\label{Eq:W}
  \end{eqnarray}
The expectation energy functional for the state $\ket{\Psi_{\mathrm{BS}}(\mathbf{r})} $ is,
\begin{eqnarray}
\bra{\Psi_{\mathrm{BS}}(\mathbf{r})}\widehat{{\cal{H}}}^{\rm{eff}} \ket{\Psi_{\mathrm{BS}}(\mathbf{r})} 
=&&\left(\bra{{\psi}_{\mathrm{vac}}(\mathbf{r})}
\displaystyle \prod_{m=1}^N \hat{c}_m 
\right) 
\sum_{i,j} \hat{c}_i^\dagger T_{ij} \hat{c}_j 
\left(
\displaystyle \prod_{m'=1}^N \hat{c}_{m'}^\dagger 
\ket{{\psi}_{\mathrm{vac}}(\mathbf{r})}\right)
\nonumber \\
&&\ \ \ \ \ \ \ \ \ \ \ \ \ \ \ \ \ \ \ \ \ \ \ 
+ \left( \bra{{\psi}_{\mathrm{vac}}(\mathbf{r})}
\displaystyle \prod_{m=1}^N \hat{c}_m 
\right)\frac{1}{2}\!\! \sum_{\substack{i,i',j,j'}} \hat{c}_i^\dagger \hat{c}_{i'}^\dagger W_{i,i',j,j'} \hat{c}_j \hat{c}_{j'} 
\left(
\displaystyle \prod_{m'=1}^N \hat{c}_{m'}^\dagger  \ket{{\psi}_{\mathrm{vac}}(\mathbf{r})}\right) \ .
\label{Eq:exp_GS}
\end{eqnarray}

Note that in the interaction term of Eq.\ \eqref{Eq:exp_GS} the contributions in the summation 
$j\!=\!j'$ and $i\!=\!i'$ will give zero.
This is because for the state $\ket{\Psi_{\mathrm{BS}}}$ there is only one particle on each site $j$ , so   $\left(\hat{c}_j\right)^2 \ket{\Psi_{\mathrm{BS}}} =0$ and $\bra{\Psi_{\mathrm{BS}}} \left(\hat{c}_j^\dagger \right)^2\!=0$.
We will see that this property is crucial for obtaining the boson solid.   Then,
\begin{eqnarray}
&&\bra{\Psi_{\mathrm{BS}}(\mathbf{r})}\widehat{{\cal{H}}}^{\rm{eff}} \ket{\Psi_{\mathrm{BS}}(\mathbf{r})} 
\nonumber \\
&&
= \sum_{i,j} \left[ \bra{{\psi}_{\mathrm{vac}}(\mathbf{r})}
\displaystyle \prod_{\substack{m=1\\ m\neq i}}^N \hat{c}_m 
\right] 
  T_{ij} 
\left[
\displaystyle \prod_{\substack{m'=1\\ m'\neq j}}^N \hat{c}_{m'}^\dagger 
\ket{{\psi}_{\mathrm{vac}}(\mathbf{r})}\right] 
+\!\frac{1}{2}\!\! \sum_{\substack{i,i',j,j'\\ i\neq i',j\neq j'}} 
\left[ \bra{{\psi}_{\mathrm{vac}}(\mathbf{r})}
\displaystyle \prod_{\substack{m=1\\ m\neq i,i'}}^N \hat{c}_m 
\right]
 W_{i,i',j,j'} 
\left[
\displaystyle \prod_{\substack{m'=1\\ m'\neq j,j'}}^N \hat{c}_{m'}^\dagger 
\ket{{\psi}_{\mathrm{vac}}(\mathbf{r})}\right]
\nonumber \\
\nonumber \\
&& =
\sum_{i,j}  T_{ij} \delta_{ij}
+\frac{1}{2}\!\! \sum_{\substack{i,i',j,j'\\  i\neq i',j\neq j'}}  W_{i,i',j,j'} 
\left( \delta_{ij}\delta_{i'j'} + \delta_{ij'}\delta_{i'j} \right)
\nonumber \\
\nonumber \\
&& =
\sum_{i} \int d^2\mathbf{r}\, \phi_i^\star(\mathbf{r}) \left(-\frac{\hbar^2\nabla^2}{2M_{\text{X}}} \right)\phi_i(\mathbf{r})
\nonumber \\
&& \ \ \ \ \ \ \ \ \ \ \ \ \ \ \ \ \ \ \ \ \ \ \ \ \ \ \  + \frac{1}{2} \sum_{\substack{j,j'\\ j\neq j'}}
 \iint  d^2\mathbf{r}\, d^2\mathbf{r}' \left( \phi_{j}^\star(\mathbf{r}) \phi_{j'}^\star(\mathbf{r}') 
t_{XX}(|\mathbf{r}'\!\!-\!\mathbf{r}|) 
\phi_j(\mathbf{r}') \phi_{j'}(\mathbf{r}) 
+\phi_{j'}^\star(\mathbf{r}) \phi_{j}^\star(\mathbf{r}') 
t_{XX}(|\mathbf{r}'\!\!-\!\mathbf{r}|) 
\phi_j(\mathbf{r}') \phi_{j'}(\mathbf{r}) \right)
\label{Eq:expGS1}
\end{eqnarray}
Using completeness  $\sum_j \phi_j^\star(\mathbf{r}) \phi_j(\mathbf{r}')\! =\! \delta(\mathbf{r}-\mathbf{r}')$, we recognize that, since  $t_{\text{XX}}(|\mathbf{r'}-\mathbf{r}|\!=\!0)=0$, the first interaction term in  Eq.\ \eqref{Eq:expGS1} is zero.  This leaves,  
\begin{eqnarray}
\bra{\Psi_{\mathrm{BS}}(\mathbf{r})}\widehat{{\cal{H}}}^{\rm{eff}} \ket{\Psi_{\mathrm{BS}}(\mathbf{r})} =
\sum_{i} \int\,d^2\mathbf{r}\, \phi_i^\star(\mathbf{r}) \left(-\frac{\hbar^2\nabla^2}{2M_{\text{X}}} \right)\phi_i(\mathbf{r})
 +\frac{1}{2} \sum_{\substack{j,j'\\ j\neq j'}} \iint\!\!  d^2\mathbf{r}\, d^2\mathbf{r}'  |\phi_{j'}(\mathbf{r})|^2  
t_{XX}(|\mathbf{r}'\!\!-\!\mathbf{r}|) |\phi_j(\mathbf{r}') |^2 \ .
 \label{Eq:expGS2} 
\end{eqnarray}

The term $j=j'$ that is absent from the sum in Eq.\ \eqref{Eq:expGS2} can alternatively be accounted for by the subtraction of a potential from an unrestricted sum over $i$ and $j$.  
Then Eq.\ \eqref{Eq:exp_GS} becomes,
\begin{eqnarray}
E(\phi_i,\phi_i^*)=
\sum_{i} \int\,d^2\mathbf{r}\, \phi_i^\star(\mathbf{r}) \left(-\frac{\hbar^2\nabla^2}{2M_{\text{X}}} \right)\phi_i(\mathbf{r})
 &+&\frac{1}{2} \sum_{\substack{i,j}} \iint\!\!  d^2\mathbf{r}\, d^2\mathbf{r}'  |\phi_{i}(\mathbf{r})|^2  
t_{XX}(|\mathbf{r}'\!\!-\!\mathbf{r}|) |\phi_j(\mathbf{r}') |^2 \nonumber \\
&-&\frac{1}{2} \sum_{i} \iint\!\!  d^2\mathbf{r}\, d^2\mathbf{r}'  |\phi_{i}(\mathbf{r})|^2  
t_{XX}(|\mathbf{r}'\!\!-\!\mathbf{r}|) |\phi_i(\mathbf{r}') |^2 \ .
 \label{Eq:expGS3} 
\end{eqnarray}

We make use of a variational approach to obtain the ground state solution  \cite{RogelSalazar2013}, 
Eq.\ \eqref{Eq:expGS3} is minimized under variations $\delta \phi_i^\star(\mathbf{r})$.  
A Lagrange multiplier $\mu$ ensures the total number of particles is conserved, $\int d^2\mathbf{r} |\phi_i(\mathbf{r})|^2 = 1$,
\begin{equation}
\frac{\delta}{\delta \phi_i^*} \left( E(\phi_i,\phi_i^*) - \mu \sum_i \int d^2\mathbf{r} |\phi_i(\mathbf{r})|^2 \right) = 0 \ .
\label{minimise}
\end{equation}
Equation \eqref{minimise} gives,
\begin{eqnarray}
-\frac{\hbar^2\nabla^2}{2M_{\text{X}}} \phi_i(\mathbf{r})
 &+&  \int\!\!  d^2\mathbf{r}'  \phi_{i}(\mathbf{r})  
t_{XX}(|\mathbf{r}'\!\!-\!\mathbf{r}|) \sum_{\substack{j}} |\phi_j(\mathbf{r}') |^2 - \int\!\!   d^2\mathbf{r}' \phi_{i}(\mathbf{r}) 
t_{XX}(|\mathbf{r}'\!\!-\!\mathbf{r}|) |\phi_i(\mathbf{r}') |^2 =\mu \phi_i(\mathbf{r}) \ .
 \label{Eq:minE_i} 
\end{eqnarray}

We turn now to the possibility of solidification coexisting with quantum coherence in a supersolid.
Equation \eqref{Eq:GroundState} for the many-body wave function of the exciton normal state solid  
consists of a product of different single-particle wave functions, each to describe an exciton on its site. 
In contrast, the many-body wave function for the exciton supersolid consists of a single function, the order parameter. The order parameter must possess off-diagonal long-range order, depending on a finite overlap between adjacent sites to establish phase coherence, and it must also possess diagonal long-range order due to the lattice symmetry. 
We propose the following form for the order parameter, 
\begin{equation}
\Psi(\mathbf{r}) = \sqrt{\rho_B} + \sum_{i=1}^{N} \phi_i(\mathbf{r}-\mathbf{r}_i) \ .
\label{Eq:OPSS}
\end{equation}
for $N$ sites with $N$ particles.
Each $\phi_i(\mathbf{r}-\mathbf{r}_i)$ is the same function but localized on different sites $i$ and centered at $\mathbf{r}_i$. 
Since $\Psi(\mathbf{r})$ has a maximum at every lattice site $i$ it possesses the diagonal long-range order of the lattice symmetry. 
The other component $\sqrt{\rho_B}$ is uniform throughout the sample and ensures the phase locking between adjacent sites made possible from the mutual overlap of the sites. This is essential for the existence of a supersolid  \cite{Leggett1970}. 

The combination of off-diagonal and diagonal long-range order makes Eq.\ \eqref{Eq:OPSS}  suitable as the order parameter for a supersolid.

Summing Eq.\ \eqref{Eq:minE_i} over the sites $i$,
\begin{equation}
 -\frac{\hbar^2\nabla^2}{2M_{\text{X}}} \sum_i \phi_i(\mathbf{r}) 
 + \sum_i \phi_i(\mathbf{r}) \! \int\!\!  d^2\mathbf{r}'   
t_{XX}(|\mathbf{r}'\!\!-\!\mathbf{r}|) \sum_{\substack{j}} |\phi_j(\mathbf{r}') |^2 
- \sum_i \phi_{i}(\mathbf{r}) \! \int\!\!   d^2\mathbf{r}'  t_{XX}(|\mathbf{r}'\!\!-\!\mathbf{r}|) |\phi_i(\mathbf{r}') |^2 = \mu \! \sum_i \phi_i(\mathbf{r}) 
\end{equation}
When the function $\phi_i(\mathbf{r}-\mathbf{r}_{i})$ is localized within a single unit cell $i$, 
\begin{equation}
 -\frac{\hbar^2\nabla^2}{2M_{\text{X}}} \sum_i \phi_i(\mathbf{r}) 
 + \sum_i \phi_i(\mathbf{r}) \! \int\!\!  d^2\mathbf{r}'   
t_{XX}(|\mathbf{r}'\!\!-\!\mathbf{r}|) \sum_{\substack{j}} |\phi_j(\mathbf{r}') |^2 
- \sum_i \phi_{i}(\mathbf{r}) \! \int_{\mathbf{r'} \in {i_\mathbf{r}}}\!\!\!\!   d^2\mathbf{r}'  t_{XX}(|\mathbf{r}'\!\!-\!\mathbf{r}|) \sum_j |\phi_j(\mathbf{r}') |^2 = \mu \! \sum_i \phi_i(\mathbf{r}) 
\label{Eq:GPstep1}
\end{equation}
and from Eq.\ \eqref{Eq:OPSS}, 
\begin{eqnarray}
|\Psi(\mathbf{r})|^2
&=& \rho_B
+ 2 \sqrt{\rho_B}\sum_i \phi_i(\mathbf{r}) 
+  \sum_i \phi_i(\mathbf{r})  \sum_{i'} \phi_{i'}(\mathbf{r}) = 
\rho_B
+ 2 \sqrt{\rho_B}\sum_i \phi_i(\mathbf{r}) 
+  \sum_i \phi_i^2(\mathbf{r})\ .
\label{Eq:Psi^2_1}
\end{eqnarray}
Equation \eqref{Eq:GPstep1} becomes,
\begin{equation}
-\frac{\hbar^2\nabla^2}{2M_{\text{X}}} \Psi(\mathbf{r})
 +  (\Psi(\mathbf{r})\!-\!\!\sqrt{\rho_B}) \!\int\!\!  d^2\mathbf{r}'   
t_{XX}(|\mathbf{r}'\!\!-\!\mathbf{r}|) \sum_{\substack{j}} |\phi_j(\mathbf{r}') |^2 
- (\Psi(\mathbf{r})\!-\!\!\sqrt{\rho_B})\int_{\mathbf{r'} \in {i_\mathbf{r}}}\!\!\!\!   d^2\mathbf{r}'  t_{XX}(|\mathbf{r}'\!\!-\!\mathbf{r}|) \sum_j |\phi_j(\mathbf{r}') |^2 = \mu (\Psi(\mathbf{r})\!-\!\!\sqrt{\rho_B}) 
 \label{Eq:minE_sumi} 
\end{equation}
The second term in Eq.\ \eqref{Eq:minE_sumi} can be written,
\begin{eqnarray}
&&(\Psi(\mathbf{r})\!-\!\sqrt{\rho_B}) \!\int\!\!  d^2\mathbf{r}'   t_{XX}(|\mathbf{r}'\!\!-\!\mathbf{r}|) \sum_{\substack{j}} |\phi_j(\mathbf{r}') |^2 
\nonumber\\
&=& \Psi(\mathbf{r})\!\int\!\!  d^2\mathbf{r}' t_{XX}(|\mathbf{r}'\!\!-\!\mathbf{r}|) \sum_{\substack{j}} |\phi_j(\mathbf{r}') |^2 -\sqrt{\rho_B} \!\int\!\!  d^2\mathbf{r}' t_{XX}(|\mathbf{r}'\!\!-\!\mathbf{r}|) \sum_{\substack{j}} |\phi_j(\mathbf{r}') |^2 
\nonumber \\
\ \\
&=&\Psi(\mathbf{r}) \!\int\!\!  d^2\mathbf{r}' t_{XX}(|\mathbf{r}'\!\!-\!\mathbf{r}|) |\Psi(\mathbf{r'})|^2 - 
\Psi(\mathbf{r}) \!\int\!\!  d^2\mathbf{r}' t_{XX}(|\mathbf{r}'\!\!-\!\mathbf{r}|)  (\rho_B +2 \sqrt{\rho_B}\sum_j \phi_j(\mathbf{r'}) )
-\sqrt{\rho_B} \!\int\!\!  d^2\mathbf{r}' t_{XX}(|\mathbf{r}'\!\!-\!\mathbf{r}|) \sum_{\substack{j}} |\phi_j(\mathbf{r}') |^2
\nonumber \\
\\
&=&\Psi(\mathbf{r}) \!\int\!\!  d^2\mathbf{r}' t_{XX}(|\mathbf{r}'\!\!-\!\mathbf{r}|) |\Psi(\mathbf{r'})|^2 
- \sqrt{\rho_B} \!\int\!\!  d^2\mathbf{r}' t_{XX}(|\mathbf{r}'\!\!-\!\mathbf{r}|)  (\rho_B +2 \sqrt{\rho_B}\sum_j \phi_j(\mathbf{r'}) ) 
\nonumber\\
&& \qquad \qquad\qquad\qquad\qquad\quad  -\sum_{\substack{i}} \phi_i(\mathbf{r})  \!\int\!\!  d^2\mathbf{r}' t_{XX}(|\mathbf{r}'\!\!-\!\mathbf{r}|)  (\rho_B +2 \sqrt{\rho_B}\sum_j \phi_j(\mathbf{r'}) ) 
-\sqrt{\rho_B} \!\int\!\!  d^2\mathbf{r}' t_{XX}(|\mathbf{r}'\!\!-\!\mathbf{r}|) \sum_{\substack{j}} |\phi_j(\mathbf{r}') |^2
\nonumber \\
\ \\
&=&\Psi(\mathbf{r}) \!\int\!\!  d^2\mathbf{r}' t_{XX}(|\mathbf{r}'\!\!-\!\mathbf{r}|) |\Psi(\mathbf{r'})|^2 
- \sqrt{\rho_B} \!\int\!\!  d^2\mathbf{r}' t_{XX}(|\mathbf{r}'\!\!-\!\mathbf{r}|)\rho_B  
-\sqrt{\rho_B} \!\int\!\!  d^2\mathbf{r}' t_{XX}(|\mathbf{r}'\!\!-\!\mathbf{r}|) (2 \sqrt{\rho_B}\sum_j \phi_j(\mathbf{r'}) +\sum_{\substack{j}} |\phi_j(\mathbf{r}') |^2)
\nonumber\\
&& \qquad \qquad\qquad\qquad\qquad \qquad\quad -\sum_{\substack{i}} \phi_i(\mathbf{r})  \!\int\!\!  d^2\mathbf{r}' t_{XX}(|\mathbf{r}'\!\!-\!\mathbf{r}|)  (\rho_B +2 \sqrt{\rho_B}\sum_j \phi_j(\mathbf{r'}) ) 
\label{Eq:minE_2nd} 
\end{eqnarray}

The third term  in Eq.\ \eqref{Eq:minE_sumi} can be written,
\begin{eqnarray}
&&- (\Psi(\mathbf{r})\!-\!\!\sqrt{\rho_B})\int_{\mathbf{r'} \in {i_\mathbf{r}}}\!\!\!\!   d^2\mathbf{r}'  t_{XX}(|\mathbf{r}'\!\!-\!\mathbf{r}|) \sum_j |\phi_j(\mathbf{r}') |^2\\
&=& -(\Psi(\mathbf{r})\!-\!\!\sqrt{\rho_B})\int_{\mathbf{r'} \in {i_\mathbf{r}}}\!\!\!\!   d^2\mathbf{r}'  t_{XX}(|\mathbf{r}'\!\!-\!\mathbf{r}|) \left[|\Psi(\mathbf{r'})|^2 - \rho_B - 2 \sqrt{\rho_B}\sum_j \phi_j(\mathbf{r'}) \right]\\
&=&-(\Psi(\mathbf{r})\!-\!\!\sqrt{\rho_B})\int_{\mathbf{r'} \in {i_\mathbf{r}}}\!\!\!\!   d^2\mathbf{r}'  t_{XX}(|\mathbf{r}'\!\!-\!\mathbf{r}|) |\Psi(\mathbf{r'})|^2 - (\Psi(\mathbf{r})\!-\!\!\sqrt{\rho_B})\int_{\mathbf{r'} \in {i_\mathbf{r}}}\!\!\!\!   d^2\mathbf{r}'  t_{XX}(|\mathbf{r}'\!\!-\!\mathbf{r}|) (\rho_B + 2 \sqrt{\rho_B}\sum_j \phi_j(\mathbf{r'}))\\
&=&-(\Psi(\mathbf{r})\!-\!\!\sqrt{\rho_B})\int_{\mathbf{r'} \in {i_\mathbf{r}}}\!\!\!\!   d^2\mathbf{r}'  t_{XX}(|\mathbf{r}'\!\!-\!\mathbf{r}|) |\Psi(\mathbf{r'})|^2 - \sum_i \phi_i(\mathbf{r})\int_{\mathbf{r'} \in {i_\mathbf{r}}}\!\!\!\!   d^2\mathbf{r}'  t_{XX}(|\mathbf{r}'\!\!-\!\mathbf{r}|) (\rho_B + 2 \sqrt{\rho_B}\sum_j \phi_j(\mathbf{r'}))
\label{Eq:minE_3rd}
\end{eqnarray}

Inserting Eqs.\ \eqref{Eq:minE_2nd} and \eqref{Eq:minE_3rd} into Eq.\ \eqref{Eq:minE_sumi} we obtain,
\begin{eqnarray}
&&-\frac{\hbar^2\nabla^2}{2M_{\text{X}}} \Psi(\mathbf{r})+ \Psi(\mathbf{r})   \int\!\!  d^2\mathbf{r}'  
t_{XX}(|\mathbf{r}'\!\!-\!\mathbf{r}|) 
|\Psi(\mathbf{r}')|^2 - \sqrt{\rho_B}\rho_B \!\int\!\!  d^2\mathbf{r}' t_{XX}(|\mathbf{r}'\!\!-\!\mathbf{r}|)
-  (\Psi(\mathbf{r})\!-\!\!\sqrt{\rho_B})\!\!\int_{\mathbf{r'} \in {j_\mathbf{r}}}\!\! \!\!\!\!d^2\mathbf{r}' 
   t_{XX}(|\mathbf{r}'\!\!-\!\mathbf{r}|)  
|\Psi(\mathbf{r}')|^2 \nonumber\\
&&\qquad\qquad\qquad\qquad\qquad\qquad\qquad\qquad\qquad\qquad\qquad\qquad + T[\rho_B, \phi_i] =\mu \Psi(\mathbf{r})-\mu \sqrt{\rho_B} \ 
 \label{Eq:GP_NSfull} 
\end{eqnarray}
where the functional $T[\rho_B, \phi_i]$ represents all the terms containing products of $\sqrt{\rho_B}$ with $\phi_j$,
\begin{eqnarray}
 T[\rho_B, \phi_i]&=&  
 - \!\int\!\!  d^2\mathbf{r}' t_{XX}(|\mathbf{r}'\!\!-\!\mathbf{r}|) \left[\sqrt{\rho_B} \sum_j \phi_j(\mathbf{r'})\right] (2 \sqrt{\rho_B} +  \sum_{\substack{j}} \phi_j(\mathbf{r}'))  \label{Eq.ExtraT}\\
&-&\sqrt{\rho_B}\sum_{\substack{i}} \phi_i(\mathbf{r})\left[ \int\!\!  d^2\mathbf{r}' t_{XX}(|\mathbf{r}'\!\!-\!\mathbf{r}|)   (\sqrt{\rho_B} +2 \sum_j \phi_j(\mathbf{r'}) ) -\int_{\mathbf{r'} \in {i_\mathbf{r}}}\!\!\!\!   d^2\mathbf{r}'  t_{XX}(|\mathbf{r}'\!\!-\!\mathbf{r}|)  (\sqrt{\rho_B} + 2 \sum_j \phi_j(\mathbf{r'}))
 \right] \nonumber
\end{eqnarray}

To obtain a unified Gross-Pitaevskii equation  for the supersolid and the superfluid, 
we recall the familiar time-independent Gross-Pitaevskii equation for a homogeneous  superfluid \cite{RogelSalazar2013},
\begin{eqnarray}
\sqrt{\rho_B}   \int\!\!  d^2\mathbf{r}'  
t_{XX}(|\mathbf{r}'\!\!-\!\mathbf{r}|) 
|\sqrt{\rho_B}|^2 
 &=&\mu \sqrt{\rho_B} \ .
 \label{Eq:GP_SF} 
\end{eqnarray}
Adding Eq.\ \eqref{Eq:GP_SF} and Eq.\ \eqref{Eq:GP_NSfull} gives the unified equation,
\begin{equation}
-\frac{\hbar^2\nabla^2}{2M_{\text{X}}} \Psi(\mathbf{r})+ \Psi(\mathbf{r})   \int\!\!  d^2\mathbf{r}'  
t_{XX}(|\mathbf{r}'\!\!-\!\mathbf{r}|) 
|\Psi(\mathbf{r}')|^2 
-  (\Psi(\mathbf{r})\!-\!\!\sqrt{\rho_B})\!\!\int_{\mathbf{r'} \in {j_\mathbf{r}}}\!\! \!\!\!\!d^2\mathbf{r}' 
   t_{XX}(|\mathbf{r}'\!\!-\!\mathbf{r}|)  
|\Psi(\mathbf{r}')|^2 + T[\rho_B, \phi_i] =\mu \Psi(\mathbf{r}) 
 \label{Eq:GP_withextra} 
\end{equation}
In the superfluid limit $\Psi(\mathbf{r})=\sqrt{\rho_B}$ and  $\phi_i(\mathbf{r})=0$.  Equation \eqref{Eq:GP_withextra} then reduces to the  Gross-Pitaevskii equation for the superfluid, Eq.\ \eqref{Eq:GP_SF}, with $T[\rho_B, \phi_i]=0$ from Eq.\ \eqref{Eq.ExtraT}. 

At the superfluid to supersolid transition $\Psi(\mathbf{r})$ will develop non-zero values of $\phi_i(\mathbf{r})$. 
If we keep $T[\rho_B, \phi_i]=0$, the calculation predicts a transition to the supersolid with $\phi_i(\mathbf{r})$ small compared to $\sqrt{\rho_B}$.
These values of $\sqrt{\rho_B}$ and $\phi_i(\mathbf{r})$ lead to very small values of $T[\rho_B, \phi_i]$ that are negligible in Eq.\  \eqref{Eq:GP_withextra}, a consistent result.

At the supersolid to normal solid transition, we expect $\sqrt{\rho_B}=0$.
If we keep $T[\rho_B, \phi_i]=0$, our calculations show that in the supersolid $\sqrt{\rho_B}$ indeed approaches zero as the interactions become very strong. 
At such small values of $\sqrt{\rho_B}$, the $T[\rho_B, \phi_i]$ determined from Eq.\  \eqref{Eq.ExtraT} is once again negligible in Eq.\ \eqref{Eq:GP_withextra}.  

We therefore propose to  interpolate between the superfluid and normal solid limits by neglecting $T[\rho_B, \phi_i]$ across the entire supersolid phase.
Our Gross-Pitaevskii equation then becomes Eqs.\ (6)-(7) in the manuscript,
\begin{eqnarray}
-\frac{\hbar^2\nabla^2}{2M_{\text{X}}} \Psi(\mathbf{r})
 &+& \Psi(\mathbf{r})   \int\!\!  d^2\mathbf{r}'  
t_{XX}(|\mathbf{r}'\!\!-\!\mathbf{r}|) 
|\Psi(\mathbf{r}')|^2 
-  (\Psi(\mathbf{r})\!-\!\!\sqrt{\rho_B})\int_{\mathbf{r'} \in {j_\mathbf{r}}}\!\! \!\!\!\!d^2\mathbf{r}' 
   t_{XX}(|\mathbf{r}'\!\!-\!\mathbf{r}|)  
|\Psi(\mathbf{r}')|^2 
 =\mu \Psi(\mathbf{r}) \ 
 \label{Eq:GP} 
\end{eqnarray}

\end{document}